\documentclass[journal=apchd5,manuscript=letter]{achemso}

\usepackage[version=3]{mhchem} 
\usepackage[T1]{fontenc}       
\usepackage{color}
\graphicspath{{figs/}}
\usepackage[export]{adjustbox}
\usepackage{natbib}
\usepackage{cuted}

\title{Holographic Resonant Laser Printing of metasurfaces using plasmonic template}

\author{Marcus S. Carstensen}
\altaffiliation{Equal contributions by X.Z. and M.S.C.}
\author{Xiaolong Zhu}
\altaffiliation{Equal contributions by X.Z. and M.S.C.}

\author{Oseze Esther Iyore}
\affiliation[Technical University of Denmark]
{DTU Nanotech, Department of Micro and Nanotechnology, Technical University of Denmark, DK-2800 Kongens Lyngby}
\affiliation[Technical University of Denmark]
{DTU Nanotech, Department of Micro and Nanotechnology, Technical University of Denmark, DK-2800 Kongens Lyngby}
\author{N.~Asger~Mortensen}
\affiliation[University of Southern Denmark]
{Center for Nano Optics \& Danish Institute for Advanced Study, University of Southern Denmark, Campusvej 55, DK-5230~Odense~M, Denmark}
\author{Uriel Levy}
\affiliation[Technical University of Denmark]
{DTU Nanotech, Department of Micro and Nanotechnology, Technical University of Denmark, DK-2800 Kongens Lyngby}
\alsoaffiliation[HUJI]
{Department of Applied Physics, The Benin School of Engineering and Computer Science, The Center for Nanoscience and Nanotechnology, The Hebrew University of Jerusalem, Jerusalem, 91904, Israel}
\author{Anders Kristensen}
\email{a.kristensen@nanotech.dtu.dk}
\affiliation[Technical University of Denmark]
{DTU Nanotech, Department of Micro and Nanotechnology, Technical University of Denmark, DK-2800 Kongens Lyngby}

\begin{document}

\begin{abstract}
Laser printing with a spatial light modulator (SLM) has several advantages over conventional raster-writing and dot-matrix display (DMD) writing: multiple pixel exposure, high power endurance and existing software for computer generated holograms (CGH). We present a technique for the design and manufacturing of plasmonic metasurfaces based on ultrafast laser printing with an SLM. As a proof of principle{\color{black},} we have used this technique to laser print a plasmonic metalens as well as high resolution plasmonic color decorations. The high throughput {\color{black}holographic resonant laser printing (HRLP)} approach enables on-demand mass-production of customized metasurfaces.
\end{abstract}

The emerging \emph{Internet of Things}\cite{buyya2016a} stimulates the development of new sensor technology, which requires cost-efficient, compact, and light-weight optical components. 
{\color{black}Such ultra-thin optical elements, of thickness comparable to the wavelength of light and even below, are achievable with optical metasurfaces:\cite{kildishev2013a,yu2014a,zhao2014a,yu2015a,Ding:2017,Hsiao:2017,Qin2016,rogers2012a,Arbabi2015,Pors2013,High2015,Khorasaninejad1190,Lin2014,Chong2015,Guo2016,Bomzon2002,Levy2004,Levy2005,Boroviks2018} Lithographically defined, spatially varying arrays of sub-wavelength dielectric or metallic elements that can control the propagation of electromagnetic radiation. In particular, metasurfaces have the capability of manipulating the phase, amplitude and polarization of light.} Recent research reports on metallic as well as dielectric or hybrid metasurfaces with diffraction-limited focusing, sub-wavelength resolution imaging and for total control of reflected or transmitted light. In this paper, we present a flexible and up-scalable method for laser printing of flat optical components in prefabricated metasurfaces, extending the concepts of plasmonic colours\cite{Kumar2012,Gu:2015a,Kristensen2016,Hedayati:2017,Duan2017} and ink-free color laser printing in metallic (plasmonic)\cite{zhu2016a,clausen2014a} and dielectric metasurfaces\cite{zhu2017a}, which can be manufactured by production-grade methods\cite{h2016a} and laser reshaping and even ablation\cite{Kuznetsov2011,Zijlstra:2009,Zhu:2017b,Makarov:2017,Gonzalez-Rubio:2017,Zuev:2016} to provide control over plasmonic colors.

In our previous work\cite{zhu2016a,zhu2017a} {\color{black}on resonant laser printing (RLP)} we obtained a world record laser printing resolution beyond {\color{black}$127,000$}\,DPI by raster scanning a focused laser beam across the metasurface to re-shape a single nano-scale metasurface element, or unit cell, at a time. As a new paradigm, holographic laser post-processing at the unit cell level is hereby introduced and used to inscribe local metasurface functionalities, {\color{black}i.e.,} beyond the control of color for {\color{black}e.g.,} high-density information storage or security marking purposes\cite{Kristensen2016}. {\color{black} In order to advance the writing speed, the laser beam is reflected on a holographic SLM (LCOS-SLM X10468) to generate and translate multiple foci ($800 \times 600$ pixels), and expose $128 \times 128$ unit cells in 100\,ms.} In this way, we improve writing speeds by  orders of magnitude.
{\color{black}SLMs} are widely used for ultra-fast 3D laser micro-machining\cite{malinauskas2016a}, known as holographic femtosecond laser processing\cite{hasegawa2014a} and laser-lithography\cite{Vizsnyiczai:14}. As a proof of concept for the effectivness of our approach, we demonstrate {\color{black}holographic resonant laser printing (HRLP)} of various flat optics components such as Fresnel zone plate (FZP) lenses with nearly {\color{black}diffraction-limited focusing} and axicons in plasmonic metasurfaces comprising a CMOS compatible approach with ultrathin aluminum (Al) films\cite{Knight:2014,Gerard:2015}. We also use the technique for single-exposure writing of plasmonic color images. Our results open a new avenue for manufacture of small series or individualized products by laser post-processing of components that are volume manufactured with a common optical metasurface template. The laser post-processing method also allows for individual alignment of optical elements on complex, or assembled components --- {\color{black}e.g.,} plastic sensor chips, as well as trimming of the optical elements even at the metasurface unit-cell level\cite{Zuev2016,Lepeshov:2017}.

Taking advantage of plasmon-enhanced light-matter interactions\cite{Chen:2012a}, we laser-post-process Al metasurfaces with morphology-dependent resonances. Strong plasmonic absorption under pulsed laser irradiation locally elevates the temperature {\color{black}in a} very short time scale (1\,ns), where rapid photo-thermal melting/sintering of the metal allows for morphology changes\cite{Chen:2012a,Zijlstra:2009,zhu2016a,Novikov:2017} with associated spatially modification of transmittance.  {\color{black}With the excitation of the surface plasmon resonances, the plasmon-enhanced photo-thermal melting ensures that the writing process only takes place at the plasmonic metasurface within the focal plane, causing a strong heat power confined at the interface and thus decreases the power consumption.} To avoid the need for tedious and time consuming scanning procedures we have developed the approach of {\color{black}HRLP}, in which an image is being projected on a uniform array of metasurfaces. It is the spatial variation of this image, generated by an SLM {\color{black}and} projected onto the metasurface plane, which controls the individual final shape of each and every metal disk, allowing the implementation of various flat optics devices and decoration effects. 

\begin{figure}
\begin{center}
\includegraphics[width=0.75\columnwidth]{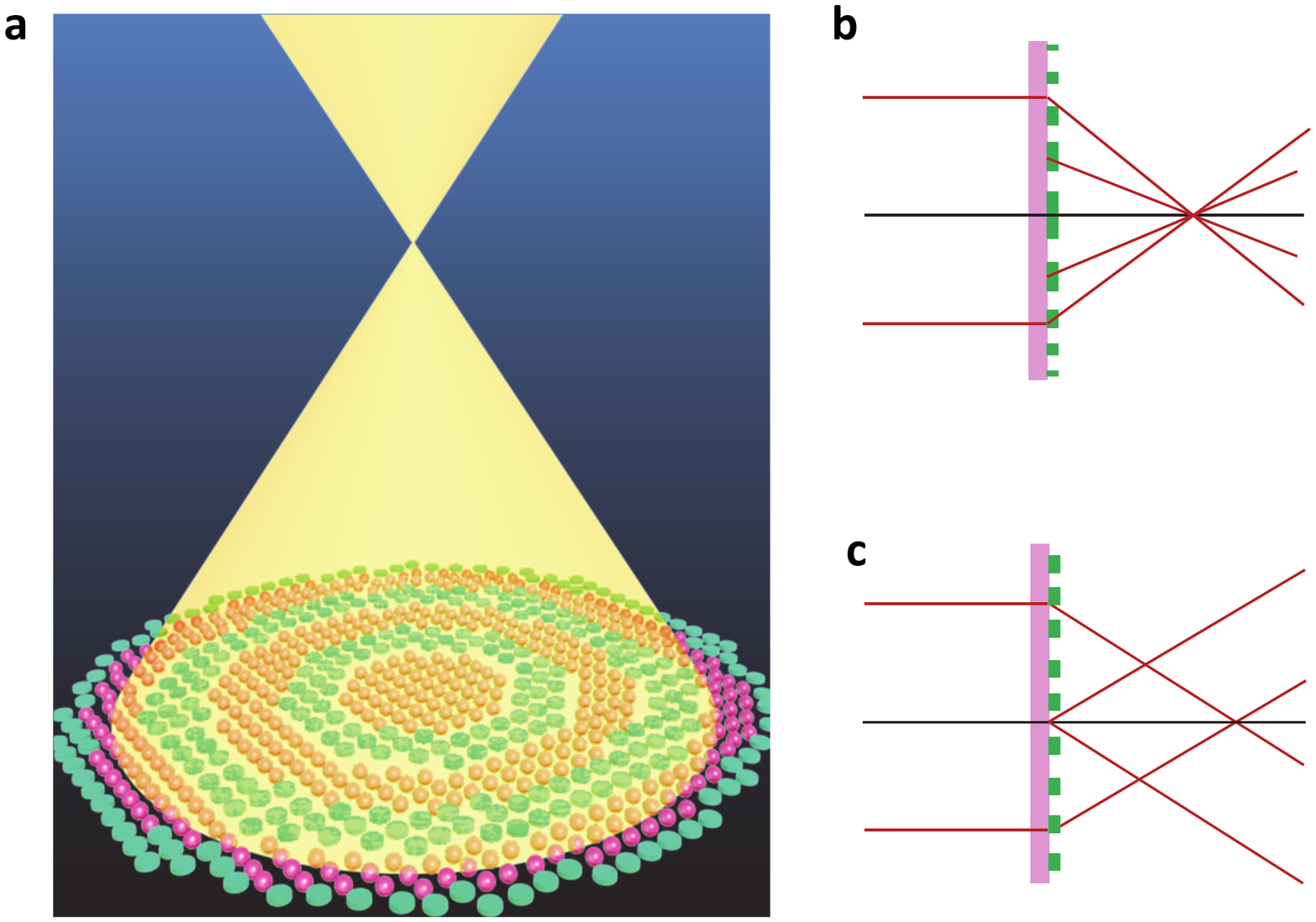}
\end{center}\linespread{1.3}
\caption{(a) Illustration of laser printed flat optics. (b) Schematics of a FZP and (c) an axicon lens made by an ultrathin patterned layer of plasmonic nanostructures.}\label{f1}
\end{figure}

The FZP\cite{Rastani:91,li2016a,wang2016a} represents one particular class of flat, thin optical lenses, where the intensity and/or the phase of the transmitted light is spatially modulated by concentric ring zones to focus the light at a given distance away {\color{black}from} the substrate. There are two classifications of FZPs, based on either amplitude or phase modulation. Clearly, one can also implement a hybrid type of FZP, where both the amplitude and the phase are spatially modulated. In all cases, the periodicity of the rings is becoming shorter towards the periphery such that the FZP supports higher diffraction angles needed for the focusing of optical rays far from the optical axis. A binary amplitude FZP, as demonstrated by Li~ \emph{et al.}\cite{li2016a}, comprises a sequence of transparent and opaque, concentric ring zones. A binary phase FZP\cite{Rastani:91} can be fabricated by etching or adding the concentric ring zones of an optically transparent material{\color{black},} whereby the optical thickness, and thereby the phase of the transmitted light is modulated spatially. The phase-based FZP provides higher diffraction efficiency. An adaptive FZP reported by Wang~\emph{et al.}\cite{wang2016a} utilizes laser-induced heating in a phase-changing material. To define the concentric ring zones, the contrast in dielectric properties is obtained by switching between an amorphous state and multiple metastable cubic crystalline states by use of a short high-density laser pulse. Hereby, we demonstrate experimentally a laser printed ultrathin FZP within plasmonic metasurfaces. Figure~\ref{f1}(a) shows a schematic of a transmissive plasmonic FZP fabricated by RLP. The building elements of the ultra-thin FZP are prefabricated plasmonic nano-resonators, which are subsequently laser re-configured. Space-variant metasurfaces can be constructed from plasmonic resonators either for focusing, Fig.~\ref{f1}(b), or for other types of beam manipulations, {\color{black}e.g.,} the construction of axicons which are typically used to form nondiffractive Bessel beams, Fig.~\ref{f1}(c). 
The {\color{black}HRLP} technology is employed to spatially modify the transmittance of the plasmonic metasurfaces. In contrast to many previous designs, our FZP is only 50 nanometer (approx. one-tenth of the wavelength) thick and can potentially be mass-produced.

\begin{figure}
\begin{center}
\includegraphics[width=1\columnwidth]{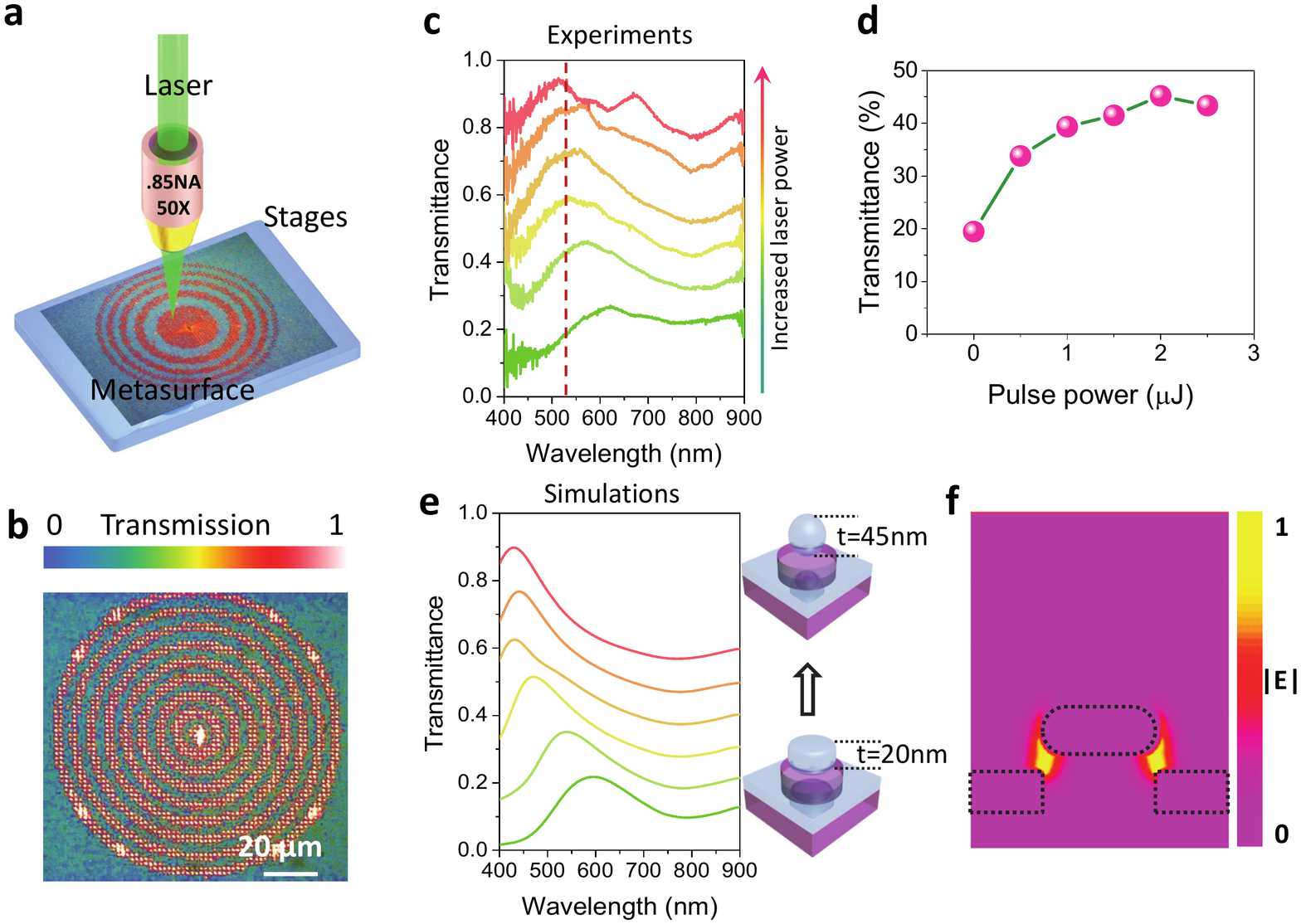}
\end{center}\linespread{1.3}
\caption{(a) schematic setup and laser printing of the ultrathin lens on optical metasurfaces. (b) Amplitude contrast presented by the transmission difference between the printed and non-printed zones. {\color{black}The transmitted signals are normalized and taken under a white-light illumination.} (c) Experimental transmittance spectra of the printed zones which are printed with different laser intensities. {\color{black}For clarity, the spectra have been stacked with a +0.1 displacement.} (d) Corresponding absolute transmittance at the wavelength of 532\,nm under different printing power which are read from (c) (indicated by the dashed red line). (e) Simulated transmittance spectra for laser modulated metasurfaces. {\color{black}For clarity, the spectra have been stacked with a +0.1 displacement.} (f) Normalized electric field distribution of the original plasmonic nanostructure at a 532\,nm excitation.}\label{f2}
\end{figure}

In plasmon-assisted laser printing, pulsed-laser irradiation generates transient thermal power in the plasmonic structures which in turn modifies the spectroscopic transmission patterns by melting and reshaping the structures. As illustrated in Fig.~\ref{f2}a, we developed a mask-free RLP technology to pattern plasmonic resonant metasurfaces with a superior resolution. This technology uses a pulsed laser (1\,ns pulse duration) with an on-resonance frequency (corresponding to a wavelength of 532\,nm) and related apparatus to control the intensity of the pulse trains, the 3D motion of the samples and {\color{black}diffraction-limited focusing} of the laser spot. As a result, different plasmonic resonances arise depending on the laser pulse energy density, which in turn leads to different transmittances as well as an on-resonance phase change. The RLP technique here was extended to an {\color{black}HRLP} technique which is developed as a flexible and single shot post-writing technology for flat optics, where rapid melting of a $\sim 100  \times 100$~${\rm \mu m^2}$ area allows for surface-energy-driven morphology changes with associated modification of amplitude, phase and polarization of the reflected, transmitted and scattered light over each individual element of the plasmonic metasurfaces. Fig.~\ref{f2}b shows the transmitted amplitude control of an axicon lens which is conducted by our {\color{black}HRLP} technique.

To manipulate the strength of the transmittance we controlled intensity of optical pulses, while preserve their repetition rate, typically at 1\,kHz. We achieved a diffraction limited resolution of printing using a lens with a numerical aperture (NA) of 0.85 and a magnification of 50$\times$ (Fig.~\ref{f2}a).
When applying {\color{black}the HRLP to} a plasmonic metasurface with a resonance located at 600\,nm, the resonant transmittance peak of the printed area blueshifts from 600\,nm to 500\,nm, which results in the contrast in the transmission images as in Fig.~\ref{f2}b. The proof of concept experiments were relayed on a plasmonic metasurface formated by depositing a thin film (20\,nm) of aluminum on top of an array of dielectric (OrmoComp, microresist technology GmbH, Berlin, Germany) pillars with a height of 30\,nm, a radius of 45\,nm and a periodicity of 200\,nm. The transmittance of the printed areas was measured by an imaging spectrometer with a grating of 300\,lines/mm (Andor Shamrock 303i and Newton 920 CCD camera). Results due to white-light illumination are shown in Fig.~\ref{f2}c. Gradually tuning the laser intensity upon printing, we demonstrated the manipulation of the transmittance at a certain wavelength ({\color{black}e.g.,} 532\,nm). As shown in Fig.~\ref{f2}d, more than 2 times transmission contrast between the pristine and printed samples can be achieved with laser pulses of a couple of $\rm\mu$Js.   

Following previous works\cite{zhu2017a,zhu2016a}, we used a simplified model of the complex thermodynamic phase transition. By sweeping the thickness (from 20\,nm to 45\,nm with a 5\,nm step, as illustrated in the right of Fig.~\ref{f2}e) of round-cornered disks (to the final spherical shape), while preserving the overall initial material volume of the disks in simulations, the plasmonic peak varies from 600\,nm to 450\,nm (Fig.~\ref{f2}e). For the transmittance, the result also matches the resonance induced enhanced transmitted signal in Fig.~\ref{f2}d. The substantial increasing of the transmittance at 532\,nm is attributed to the fact that the melted disks together with the underneath holes support a strong hybridized plasmonic resonance.\cite{clausen2014a} The excitation of that resonance arouses the extraordinary optical transmission which is now  well-known to be due to the interaction of the light with electronic resonances in the surface of the metal film,\cite{Ebbesen2007} as shown in Fig.~\ref{f2}f. 
{\color{black}It should be mentioned that we used a simplified model of the complex thermodynamic phase transition. In Fig.~\ref{f2}c, the peaks are seen in the experimental spectra for the highest laser intensities. This can be attributed to the further laser-induced modification of the structure, leading to multiple resonances as well as the extra peaks in the transmittance spectra.}

\begin{figure}
\begin{center}
\includegraphics[width=1\columnwidth]{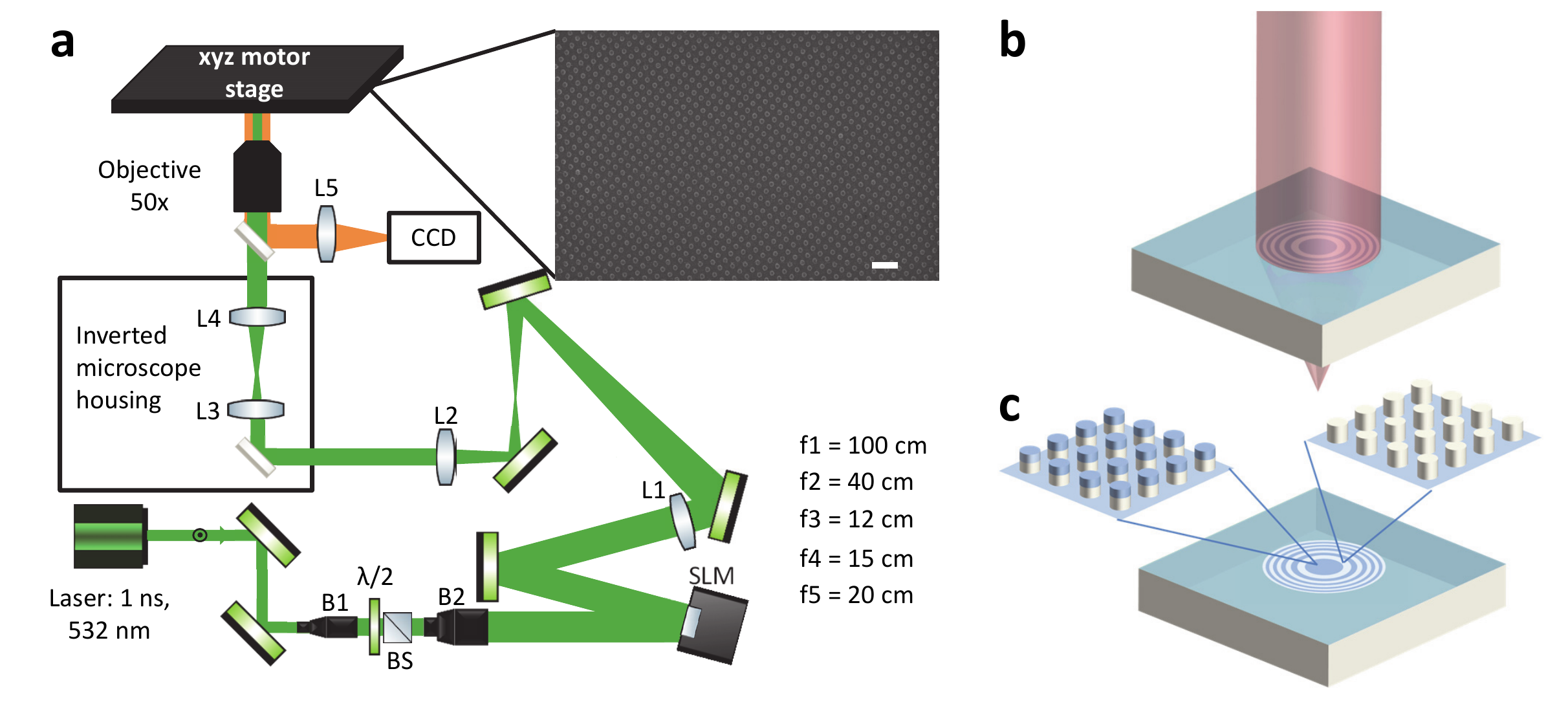}
\end{center}\linespread{1.3}
\caption{(a) Optical configuration for single-shot laser printing with the aid of a spatial light modulator (SLM). We used a half-wave plate and a beam splitter (BS) to continuously modulate the laser pulse energy when printing. Beam expanders (B1 and B2) were employed to match the spot size onto the SLM window. Several lenses (L1 to L4) are used to generate and recover the Fourier plane at the end-surfaces, as well as to ensure full coverage of the aperture of the microscope objective. Inset shows a SEM image of a representative plasmonic metasurface. Scale bar: 500\,nm. (b) and (c) Concept of laser printed flat optics, here illustrated by the writing of a Fresnel zone plate (FZP) in a plasmonic template.}\label{f3}
\end{figure}

Holographic resonant laser printing with an SLM has several advantages over conventional raster-writing\cite{Chen:2016a} and dot-matrix display (DMD) writing: multiple pixel exposure, high power endurance and existing convenience for computer generated holograms (CGH), see also the supplementary information. 
Fig.~\ref{f3}(a) shows the optical setup that has been used for the {\color{black}HRLP}. Briefly, the beam emerging from a frequency-doubled Nd:YAG laser is {\color{black}expanded}, and its intensity is controlled via a polarized beam splitter and a half waveplate. Next, light is impinging on the SLM, and the desired hologram is being created. Finally, the image of the {\color{black}hologram} is projected onto the back aperture plane of a microscope objective. In the focal plane of the microscope objective{\color{black},} a Fourier transform of the image is being created {\color{black}and} this desired light distribution is interacting {\color{black}with} the metasurface pattern. 
Fig.~\ref{f3}(b) and (c) show the concept of laser printed flat optics, illustrated by the writing of a Fresnel zone plate (FZP) in a plasmonic template constituted by \emph{hybridized nanodisk and nanohole arrays}, initially developed for plasmonic colors\cite{clausen2014a}, and now available through mass-production techniques\cite{h2016a}.
{\color{black}Because of advantages of HRLP, most of the current commercial lasers can provide the needed power even for laser printing with hi-res SLMs, for instance, the 4K SLM with $3840 \times 2160$ pixels (EXULUS-4K1, Thorlabs), which will highly improve the uniformity of exposed patterns.}

\begin{figure}
\begin{center}
\includegraphics[width=0.75\columnwidth]{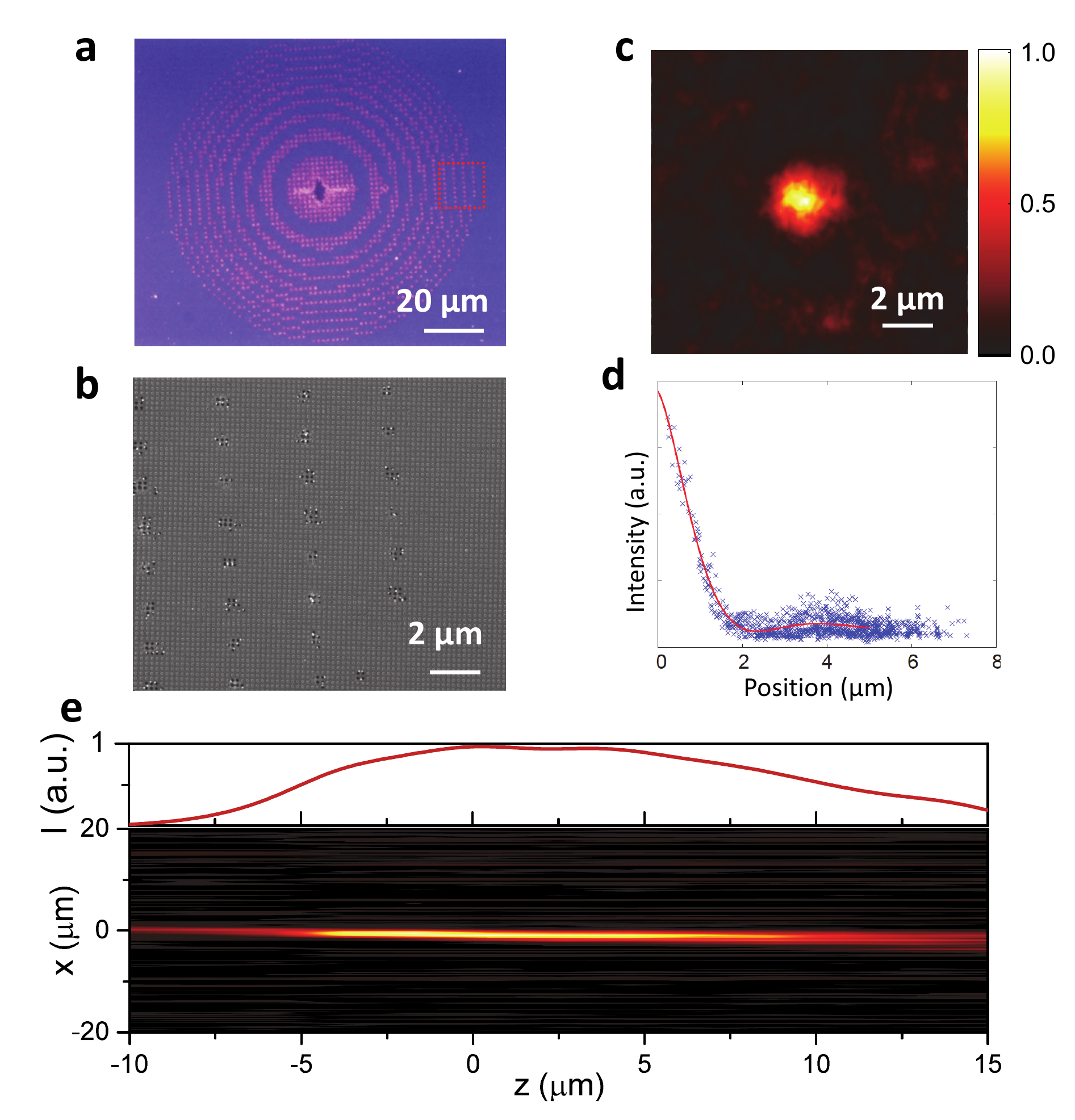}
\end{center}\linespread{1.3}
\caption{(a) Microscopic image of a fabricated FZP. Scale bar: 20\,$\rm\mu$m. (b) SEM image of a selected region (red box) in (a) which shows both the printed and unprinted areas. Scale bar: 2\,$\rm\mu$m. (c) Experimentally obtained image of the focused plane under a 532\,nm laser illumination. (d) Fitted experimental focal field intensity for laser illumination with wavelength of 532\,nm by integrating the intensity signals in a radial manner, which results in (a) . (e) Measured beam intensity profile of the FZP in the axial direction around the focal point. The intensity (I) along the center of focal beam is plotted along the $z$ axis.}\label{f4}
\end{figure}

To realize easy-to-fabricate ultra-thin flat FZPs, we implemented the {\color{black}HRLP} by an SLM providing  {\color{black}$800 \times 600$ pixels}, where the laser intensity is varying dynamically in space. Fig.~\ref{f4}a shows a printed plasmonic metasurface optical component (100\,$\rm\mu$m in width) with an optimized transmission contrast serving as a focusing lens.
While classic silica lenses are several millimeters thick, the plasmonic FZP features a 50\,nm functional layer of Al disk-hole structures. Fig.~\ref{f4}b shows a scanning electron microscopy (SEM) image of a part of the fabricated FZP with exposed and unexposed areas of outermost rings composed of dense plasmonic elements. 
{\color{black}Typical hologram from SLM achieves diffraction pattern in the first diffraction order. The remaining light intensity is mostly located in the zero-diffraction order, which results in the defects in HRLP. In order to avoid this zeroth peak from the SLM, one can adjust the focus positions and transform the original on-axis diffractive images into off-axis ones by combining a blazed grating or a FZP in the original CGH patterns. However, the method to spatially separate the undesired zeroth order from the first order has disadvantages, such as the geometric distortions, the beam aberrations and the decreasing of the printing energy. In this work, the original CGH is used as a proof-of-concept method for HRLP. Noted that the light transmitted from the defect consists a very small portion of the light at the focal point when considering the size.}

When illuminated with a coherent plane wave at a wavelength of 532\,nm, the printed FZP creates a highly symmetric focal spot at a distance of 258\,$\rm\mu$m, as shown in Fig.~\ref{f4}c. 
Note that the experimental focal distance is slightly shorter than the theoretical design (300\,$\rm\mu$m), which probably originates from limited fabrication precision while performing the pixelized Fourier transformation within {\color{black}the} SLM.
{\color{black}Fig.}~\ref{f4}d shows a diffraction-limited ($\lambda/2{\rm NA}$) full-width at half-maximum (FWHM) of about 1.5\,$\rm\mu$m by integrating the intensity signals in a radial manner of the focal spot. In addition, we also measured the beam intensity profile of the FZP in the axial direction around the focal point (Fig.~\ref{f4}e). It should be mentioned that the long-working-distance flat lens may play a role in applications such as optofluidics\cite{Psaltis:2006} or optical trapping\cite{Woerdemann:2013}. 

\begin{figure}
\begin{center}
\includegraphics[width=1\columnwidth]{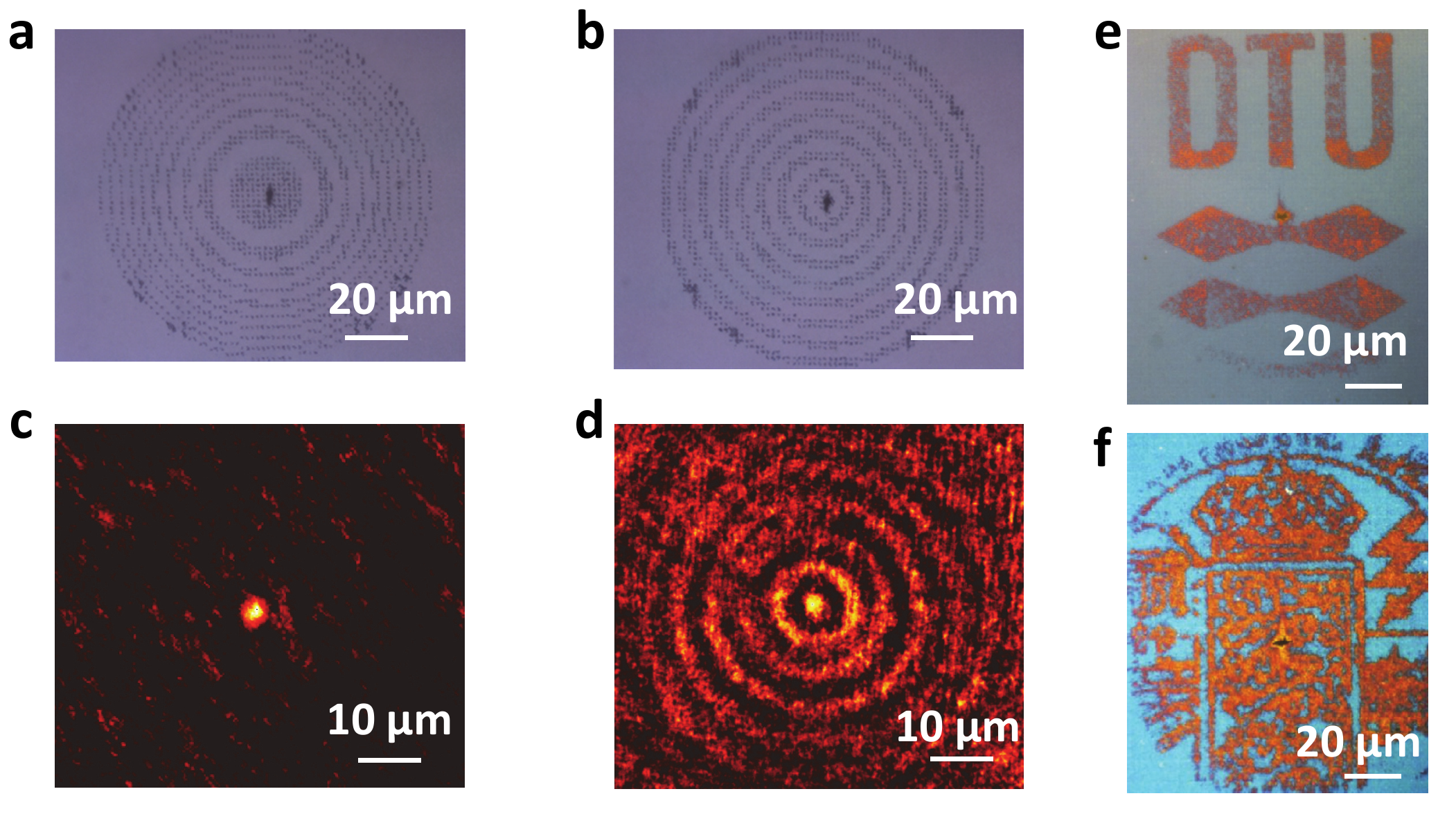}
\end{center}\linespread{1.3}
\caption{(a) Microscopic images of a fabricated FZP and (b) an axicon on plasmonic metasurface fabricated by DUV lithography. Scale bar: 20\,$\rm\mu$m. (c) Experimentally obtained
images of the focused plane via the corresponding FZP and (d) axicon under a 635\,nm laser illumination. (e,f) Graphics printed by holographic laser beam reconstructions with the SLM.}\label{f5}
\end{figure}

Because of its full flexibility, large-scale capability and direct one-step process, {\color{black}HRLP} may revolutionize the conventional product chain for optical systems and has the potential to commercialize the integrated optoelectronic system with printed flat optics. To demonstrate the diversity of our technology and its strength for end-products, we next relax
the extreme conditions of this method. Plasmonic metasurfaces made by deep-UV stepper lithography (a main-steam industrial manufacturing tool) were also fabricated and subsequently employed and HRLP-patterned with other functional flat optical components. When illuminated with a coherent plane-wave light beam at a 635\,nm wavelength, laser printed metasurface optical components serving as a lens (Fig.~\ref{f5}a, see also in the Supplementary information) and an axicon (Fig.~\ref{f5}b) create a single focal spot (Fig.~\ref{f5}c) and a nondiffractive Bessel beam (Fig.~\ref{f5}d), respectively. Moreover, the laser printing on plasmonic colored metasurfaces with spatial wave shaping by {\color{black}the} SLM is also immediately applicable for more efficient plasmonic color printing. {\color{black}Fig,~\ref{f5}e and \ref{f5}f} present graphics in red printed by holographic laser beam reconstructions. It is worthy to emphasize that the results further reveal the strength of {\color{black}HRLP} for flat optics, high definition and ink-free color printing and with a potential for future functional metasurfaces.

To sum up, as a superior alternative to using state-of-the-art and costly fabrication technologies, we demonstrate that {\color{black}HRLP,} which is realized by applying opto-thermal modification of individual nanoscale elements, combined with holographic projection of an image pattern using an SLM{\color{black},} is a powerful tool for the fabrication of ultrathin flat optics within plasmonic metasurfaces. Ultra-thin flat FZPs and axicons capable of generating diffraction-limited focal spots and nondiffractive Bessel beams are achieved with the {\color{black}HRLP} process. The concept of {\color{black}HRLP} makes the meta-optics closer to reality by providing a path for mass-production and ready-for-applications technique. This may pave the way of ultrathin flat optics into consumer products in everyday life.

\begin{acknowledgement}
This work was supported by the European Commission through the H2020 FET-OPEN project CHROMAVISION (Grant Agreement no. 665233) and H2020-NMP-PILOTS IZADI-NANO2INDUSTRY (Grant Agreement no. 686165), and by the International Network Programme of the Danish Agency for Science, Technology and Innovation (1370-00124B \& 4070-00158B). N~.A.~M. is a VILLUM Investigator (grant No. 16498) and X.~Z. is funded by VILLUM Experiment (grant No. 17400), both supported by VILLUM FONDEN. The authors thank C. Wolff for help with data processing and C.-W.~Qiu and Y.~W.~Huang for fruitful discussions.
\end{acknowledgement}

\newpage

\bibliography{Zhu}

\end{document}